\documentstyle[12pt,aaspp4,flushrt]{article}
\newcommand{\nd}{N_{\rm dens}}
\newcommand{\nh}{N_{\rm hop}}
\newcommand{\nm}{N_{\rm merge}}
\newcommand{\dout}{\delta_{\rm outer}}
\newcommand{\dm}{\delta_{\rm peak}}
\newcommand{\ds}{\delta_{\rm saddle}}
\begin{document}
\title{HOP: A New Group Finding Algorithm for N-body Simulations}
\author{Daniel J. Eisenstein and Piet Hut}
\affil{Institute for Advanced Study, Princeton, NJ 08540}
\begin{abstract}

We describe a new method (HOP) for identifying groups of particles in
N-body simulations.  Having assigned to every particle an estimate of
its local density, we associate each particle with the densest of the
$\nh$ particles nearest to it.  Repeating this process allows us to
trace a path, within the particle set itself, from each particle in the
direction of increasing density.  The path ends when it reaches a
particle that is its own densest neighbor; all particles reaching the
same such particle are identified as a group.  Combined with an adaptive
smoothing kernel for finding the densities, this method is spatially
adaptive, coordinate-free, and numerically straight-forward.  One can
proceed to process the output by truncating groups at a particular
density contour and combining groups that share a (possibly different)
density contour.  While the resulting algorithm has several user-chosen
parameters, we show that the results are insensitive to most of these,
the exception being the outer density cutoff of the groups.

\end{abstract}
\bigskip\noindent
\bigskip\bigskip
\clearpage

\section{Introduction}
Many astrophysical projects hinge upon assigning importance to
overdense regions in a set of points.  Examples include identifying
collapsed halos in a cosmological simulation, determining whether two
galaxies have merged, finding clusters of galaxies in a survey, or
locating dwarf galaxies in star counts.  Because the objects
represented by the groups of close points often have fuzzy edges that
may overlap with other groups, there is no perfect algorithmic
definition of a group; consequently, the answers to questions posed 
about the objects may depend upon the way in which membership in
the groups was decided.  It is therefore useful to have several
different methods available for dividing a set of points into groups
of associated particles; each will have its advantages and drawbacks.

In gravitational N-body simulations, one wishes to associate the clumps
of particles with real-world objects, be they galaxies, dark matter
halos, or clusters.  While one has full 6-dimensional phase space
information, in practice the particles cover phase space too sparsely
to provide unambiguous differentiation of groups on the basis of velocity.
Hence, as a first cut, it is usual to consider only the spatial
information, in which one looks for density enhancements.  We focus
here on methods that divide the set of particles into equivalence classes,
i.e.\ that each particle is a member of one and only one group;
one could imagine alternatives in which particles could belong to a
series of nested or overlapping groups or even have their group
membership expressed only in a probablistic way.

Some of the more popular group-finding algorithms are friends-of-friends
(\markcite{Dav85}Davis et al.\ 1985; see \markcite{Bar87}Barnes \&
Efstathiou 1987 for a generalization), 
DENMAX (\markcite{Ber91}Bertschinger \& Gelb 1991;
\markcite{Gel94}Gelb \& Bertschinger 1994; \markcite{Fre95}Frederic
1995), and various methods based on the overdensity within spherical
regions (\markcite{War92}Warren et al.\ 1992; \markcite{Lac94}Lacey \&
Cole 1994; \markcite{Bon96}Bond \& Myers 1996).  Because they raise
issues related to our method, we will describe the first two of these
in more detail.

In the friends-of-friends (FOF) algorithm (\markcite{Dav85}Davis et
al.\ 1985), two particles are part of the same group if their
separation is less than some chosen value; chains of such pairs then
define the groups.  The method is therefore coordinate-free and has
only one parameter; moreover, the outer boundary of the groups tend
roughly to correspond to a density contour, related to the inverse cube
of the linking distance.  Unfortunately, some groups found by the
method may appear to the eye as two clumps, linked by a small thread of
particles running between the subgroups.  This can be harmless for some
applications but may be inappropriate for others.

DENMAX takes a quite different approach, first estimating the density
at each point in space and then allowing each particle to determine its
group membership by tracing a path along the gradient of this density
surface until it reachs a local maximum.  All particles which end up in
the same maximum are assigned to the same group.  Unfortunately, the
resolution with which one defines the density field results in a
trade-off between one's ability to recognize smaller groups and one's
propensity to split groups due to the discovery of multiple local
maxima at finer resolution.  These splittings may be undesirable, and
it is difficult to assign physical importance to the way in which
low-density particles are assigned into groups on the basis of the
details of the locations of multiple local maxima in the high-density
regions.  Some implementations of DENMAX (\markcite{Gel94}Gelb \&
Bertschinger 1994) calculate the densities using a grid, although a
more recent version known as SKID (\markcite{Sta97}Stadel et
al.\ 1997; \markcite{Wei97}Weinberg et al.\ 1997) employs a
coordinate-free method.

In this Letter, we describe a new method, dubbed HOP, similar in spirit
to DENMAX but rather different in its implementation.  Instead of
constructing the density gradient field, we confine ourselves to the
point set, associate a density to every particle, and replace the
concept of the gradient with a simple search for the highest density
among a particle's nearest neighbors.  We link each particle to its
densest neighbor and then on to that particle's densest neighbor and so
on, until we reach a particle which is its own densest neighbor.  All
particles that are traced to the same such particle, hereafter called a
maximum, constitute a single group.  Because each ``hop" moves to a
higher density particle, this process is guaranteed to converge.  We
thus avoid any numerical integration of the density gradient field,
with its associated subtlety of a stopping condition; moreover, the
method is coordinate-free and can be spatially adaptive if the scales
used to determine the density and the hopping process are tuned to
reflect the local density, for example by using a constant number of
nearest neighbors.

At this point, every particle is assigned to a group; since we are 
interested in the high-density regions, we simply remove particles
below a given density threshold.  However, one now finds that the
method will have split the particles within that particular density
contour into several pieces, one for each maximum within.  As this may be
undesirable, we reconnect those groups that share a sufficiently dense
boundary, using two density thresholds to guard against small
fluctuations in a contour splitting a group in two.  Finally, we
demonstrate that, while the method thus has three integral parameters
for near-neighbor searching and three density thresholds to be chosen,
in fact the results are insensitive to one's choices for five of these:
only the outer density contrast is important for determining group properties.

\clearpage
\section{Method}
\subsection{Finding Groups by Hopping}
We wish to divide the particles into distinct sets such that particles 
in individual high-density regions are grouped together and left
separate from those in other regions.  To do this, we attempt to
distinguish between nearby groups by assigning each particle to the
group nucleus that it reaches by following in the direction of increasing
density.  Rather than construct an estimate of the density gradient at
any point in space, we instead calculate an estimate of the density at
every particle position and then determine, of the particle and its
nearest $\nh-1$ neighbors, which of the particles has the highest density.
We next associate each particle with its highest density neighbor and
continue hopping to higher and higher densities until we reach a
particle that is its own highest density neighbor (a ``maximum").
Because one always hops to increasing density, it is impossible to
enter a non-convergent loop.  All particles that hop to the same
maximum are placed into a group.  Every particle is assigned to one
and only one group; hence, we are establishing a set of equivalence
classes among the particles.

We assign a density estimate to each particle by using the radial
positions of the $\nd$ nearest neighbors of the given particle.
Generally, we use the SMOOTH algorithm (\markcite{Sta95}Stadel 1995),
which calculates the densities from the $\nd$ by a spherically
symmetric cubic spline kernel (\markcite{Mon85}Monaghan \& Lattanzio
1985), giving unit weight to particles at zero distance from the given
particle and zero weight to those at distances equal to or greater than
the radial distance to the $\nd$ nearest neighbor.  We also ran some
tests with the simpler prescription that the density is proportional to
$(3/4\pi)\nd r^{-3}_N$ (\markcite{Cas85}Casertano \& Hut 1985).  In
either case, the smoothing length for calculating the density
automatically adapts to the density itself, becoming smaller in
high-density regions.

As such, the above algorithm for assigning particles to groups is
spatially adaptive, coordinate-free, and involves two parameters, $\nh$
and $\nd$.  Once the densities are found, the group-finding proceeds
without the numerical subtleties that DENMAX must face; there are no
differential equations to solve and the stopping condition is trivial.
In general, quantities other than the density, e.g.\ the gravitational
potential or the magnitude of the acceleration (\markcite{Lac94}Lacey
\& Cole 1994), could be substituted.

The HOP algorithm could be useful for other applications
besides group finding in large-scale cosmological simulations.  For
example, to study the process of galaxy merging, one might set up a
series of automated scattering experiments in which each experiment
starts off with two galaxies approaching each other with different
relative velocities and orientations.  During such experiments, the HOP
algorithm could be used to flag the occurrence of a merger, from the fact
that the total number of groups has been reduced from two to one.  Of
course, additional care would have to be taken to check that the merger is
permanent, rather than a temporary stage during which the two galaxies
pass through each other only to separate again later.

\subsection{Merging Groups by Boundaries}
In many cases, we are interested simply in 
distinguishing dense clumps from less-dense regions and wish to 
ignore the substructure within highly dense regions.
In this case, the algorithm as it stands has two problems.  First, 
all particles are assigned to groups, so no distinction between 
the dense halo and its surroundings has yet been made.  Second, because
the density is determined adaptively, even highly dense regions may
have multiple maxima, causing the group to split into many pieces
with unphysical shapes, e.g.\ a spherical halo might be cut into sectors.
Here we offer solutions to these two problems.

As the density around each particle has already been computed, the obvious
solution to the first problem is to only include particles that exceed
some density threshold $\dout$.  As this places us in a position to
select groups based on density contours, we address the second problem
in the following way.  We adopt two additional density thresholds, 
$\ds$ and $\dm$, usually picked so that $\dout\le\ds\le\dm$.
We next define a boundary pair between two groups when
a particle and one of its $\nm$ nearest neighbors are in different groups;
the density of the boundary pair is defined to be the average of the densities
of the two particles.  We then merge two groups if their densest boundary pair
(which is our equivalent of a saddle point) exceeds the density 
$\ds$.  Furthermore, groups whose maximum
densities (i.e. the density of the group maximum) are less than $\dm$
are disbanded unless they share a boundary pair denser than $\dout$ with
a group whose density does exceed $\dm$; such groups are attached to
the group with which they share the largest boundary pair.

We depict these rules visually in Figure 1.
The intention of the algorithm is that groups consist
of central regions whose maximum densities exceed $\dm$ but whose
boundaries are defined by contours of a lower density $\ds$; the connectivity
of the $\ds$ contours near ``pinch points'' is adjudicated via the $\nm$
nearest neighbors.  However, from these centers, the groups extend out
to a yet smaller density contour of $\dout$, with affiliation between rival
disjoint interior contours of density $\ds$ being determined by the
process of hopping towards increasing density.  In the rarer case where
that hopping takes one to a density maximum with
$\dout<\delta<\dm$, the entire subgroup is attached to the viable
($\delta>\dm$) group with which it shares the highest density contour.

Dealing with these two problems has introduced four new
parameters---$\dout$, $\ds$, $\dm$, and $\nm$---more than one might
have wished for.  Clearly, the definition of the outer boundary of a
group must involve a parameter such as $\dout$.  We shall demonstrate
below that the results of merging are insensitive to the value of
$\nm$.  Finally, the possibility of choosing $\dm\ne\ds\ne\dout$ offers
the following advantages.  Simply selecting groups on the basis of a
single density contour (i.e.\ $\dm=\ds=\dout$) can produce groups that
appear by eye to be multiple groups linked by thin bridges.  By
allowing $\ds>\dout$ one suppresses such groups by splitting them
according to the topology of a second, more interior, contour.  By
using $\dm>\ds$ one effectively blurs out this boundary so that a group
cannot split in two due to minor fluctuations exactly at the $\ds$
contour.  For example, in Figure 1, if $\ds=\dm$, the group EFG would
have been split due to a small fluctuation in the contour; by using
$\ds\ne\dm$, this split is avoided.

\subsection{Implementation}
The HOP method relies heavily on being able to efficiently extract
lists of the $N$ nearest neighbors for each particle.  For this, we
use the tree-based near-neighbor search algorithm made publically
available as part of the SMOOTH program (\markcite{Sta95}Stadel 1995).
To implement HOP, we modified their program so as to alter how 
these lists of neighbors were used.

We make three passes through the data set.  First, we calculate the
density around each particle using $\nd$ nearest neighbors.  Second, we
search for the densest of each particle's $\nh$ nearest neighbors.
After tracing each particle to its maximum in order to label the
groups, we make a third pass with $\nm$ neighbors, compiling the
densest boundary pairs between each pair of groups.  We then write as
output the densities and group memberships of each particle as well as
all the densest boundary pairs.  We can then quickly and repeatedly apply 
different choices for the density cut ($\dout$) and group merging
parameters ($\ds$, $\dm$), allowing us to vary these thresholds
without rerunning the main code.

For our tests, we use a single time output of a cold dark matter
simulation, kindly provided by Guohong Xu, Renyue Cen, and Jeremiah
Ostriker.  The simulation, run with the Tree Particle Mesh (TPM) method
(\markcite{Xu95}Xu 1995), has 16.8 million particles and the non-linear
mass scale (where the top-hat {\it rms} mass fluctuations equal 1.69)
contains about 12,000 particles.  The HOP code runs on this simulation
in somewhat under 2 CPU hours on a UltraSparc 170E (for $\nd=64$,
$\nh=16$).  More than half of this time is spent calculating the
densities, as this requires the largest number of neighbors to be
found.

\subsection{Insensitivity to Parameters}
With six free parameters, it is important to test the sensitivity of
the results to changes in these parameters.  As stated above, it is
clear that at least one of these parameters, the density threshold used
to clip the outer extent of the groups, will make a significant
difference.  We will argue, however, that varying the remaining five
parameters produces little change in the results.

We adopt as our canonical parameter set $\nd=64$, $\nh=16$, $\nm=4$,
$\dm=160$, $\ds=140$, and $\dout=80$, with densities normalized
relative to the mean density.  We adopt the notation
($\nd$--$\nh$--$\nm$)[$\dm$--$\ds$--$\dout$] to label sets.  In Figure
2, we present the fraction of the total mass held in groups per
logarithmic interval of group mass (i.e.\ membership number).  The
effects of varying $\nh$ from 8 to 64 are shown; the differences are
small, but as expected, smaller $\nh$ produce slightly more small
groups, while larger $\nh$ cause some of the small groups to be
swallowed by larger groups.

In Figure 3, we show the same statistic, but we vary $\nm$.  There is
practically no difference between $\nm=4$ and $\nm=8$.  $\nm=1$ allows
slightly more small groups to survive unmerged.

In Figure 4, we show the same statistic while comparing $\nd$
of 64 and 128.  The most important result here is that groups below the
scale $\nd$ tend not to be identified.  Presumably clumps of only a few
particles either do not achieve sufficient density after smoothing to be
included or, if they do get included, get absorbed into larger groups.
Also, for $\nd=128$, about 4\% fewer particles have sufficiently high
density $\dout$ (80 in this case) to be in any group; most of this
discrepancy is on the small mass end.  We also show results when
estimating the density as the inverse cube of the distance to the 24th
nearest neighbor (i.e.\ using a top-hat kernel rather than a cubic spline).
This slightly reduces the size of large groups while identifying more
small ones.

In Figure 5, we show the same statistic while varying $\dm$ and $\ds$ at fixed
$\dout=80$.  We set $\ds$ to be $(3\dm+\dout)/4$.  As $\dm$ increases,
group sizes decrease as fringe groups are left unattached to large
groups.  Also, the total number of particles in groups decreases
because more and more groups are disqualified for failing to have 
a maximum density
in excess of $\dm$.  This is particularly important for the smallest
groups, in that it is nearly impossible for a group to have one
particle with a density of, say, 800, while having only 100 particles
above a density of 80.  We prefer $\dm$ to be a small multiple of
$\dout$, perhaps 2 or 3.  

In Figure 6, we show the same statistic, fixing $\dm=240$ and $\dout=80$ and
letting $\ds$ vary between the two.  For $\ds=\dout$, the method
selects groups as any closed contour of density $\dout$ that achieves a
density $\dm$ somewhere within it.  As $\ds$ increases, a group may be
subdivided as the region interior to contours of density $\ds$ goes
from one connected volume (at $\dout$) to multiple disjoint volumes.
At $\ds=\dm$ each closed contour of density $\dm$ defines a group
center, with the group membership extending out to $\dout$ by hopping.
Hence, as $\ds$ increases, groups are broken into pieces, producing
fewer large groups and more small groups, as seen in the figure.

In Figure 7, we show the same statistic, varying $\dm$, $\ds$, and
$\dout$ but keeping the ratio fixed.  We also show the results for two
FOF groupings, with link lengths of 0.1 and 0.2 in units of the mean
particle separation (denoted FOF(0.1), etc.).  At certain choices of
$\dout$, the results of our method produce a rather similar plot to
those of FOF, suggesting that both algorithms are approximating
criteria based on density contours.  As discussed above, the existence
of a second density parameter $\dm$ in HOP produces slight differences
from FOF in the case of groups with multiple centers.  Also note that
the two methods produce very different behavior for groups with fewer
than 50 to 100 particles.  In particular, FOF finds a very large number
of small groups, mostly in the fringes of larger groups
(\markcite{Bon91}Bond et al.\ 1991).  The shapes of the FOF curves in
the plot show that this behavior is simply a numerical artifact---there
is no physical effect to provide the upturn at $N\approx50$.  This
reinforces the lesson that the identification of small groups, even up
to dozens of particles, may not be trustworthy when the groups are well
below the non-linear mass scale and selected only with the particle
position data.  In contrast, HOP produces almost no small groups.  Of
course, this is also a numerical artifact; since the density is found
using the nearest $\nd$ particles, it is difficult for a group to have
a maximum density exceeding $\dm$ (as required to be viable) and yet
have many fewer than $\nd$ particles with density above $\dout$.

\section{Discussion}

At heart, our method relies on the idea that searching among the
nearest $N$ particles is an efficient way to collect particles into
density maxima.  In this sense, the HOP algorithm is an attractive
alternative to the usual numerical treatments of integrating the
gradient of an interpolated density field.  By choosing different means
of estimating the density, or even choosing other scalar fields to
maximize, one can imagine a variety of different group-finders.  For
example, by estimating the density with a fixed smoothing scale, one
would stay more faithful to the original DENMAX scheme.

In our particular implementation, we use an adaptive smoothing scale
for the density estimates.  By avoiding a fixed smoothing length, we
retain sensitivity to smaller groups.  However, the resultant small
smoothing lengths in high density regions can cause groups to be split
into multiple pieces due to the presence of several local maxima in
their centers.  These pieces generally do not represent true
substructure, so we wish to recombine them.  We do this by merging
groups that share high-density boundaries.  The two density scales used
for this may be chosen to be different from the density contour that
defines the outer boundary of the group(s), so as to guard against the
merging of groups connected only by thin necks.  To quantify this, we
consider the ratio of the two largest eigenvalues of the moment of
inertia tensor of groups with more than 1000 particles and counted the
fraction of such groups with a ratio greater than 2.5.  We found that
4.6\% of FOF(0.2) groups were this prolate, while for HOP with
($\nd$--$\nh$--$\nm$) = (64--16--4) and [$\dm$--$\ds$--$\dout$] =
[80--80--80], [160--140--80], [240--200--80], and [800--620--80] the
fractions were 6.6\%, 3.1\%, 1.8\%, and 0.9\% respectively.  Hence, a
ratio of $\dm$ to $\dout$ of 2--3 substantially reduces the number of
extremely prolate objects.  Perhaps extensions of FOF using the
generalization of Barnes \& Efstathiou (\markcite{Bar87}1987) or
techniques of minimal spanning trees (\markcite{Bar85}Barrow et
al.\ 1985; \markcite{Bha96}Bhavsar \& Splinter 1996) could provide
other ways to suppress these thin-neck groups.

The HOP algorithm has six free parameters; however, we have
demonstrated that the results are quite insensitive to all but one:
the choice of outer density contour $\dout$.  We recommend a default choice
of (64--16--4)[3$\dout$--2.5$\dout$--$\dout$].  The resulting behavior
is rather similar to the friends-of-friends algorithm, but with the
advantages that group membership is explicitly based on the smoothed
density field rather than pair percolation and that groups prematurely
merged by thin necks may be separated by use of a second density
threshold.

We thank Renyue Cen, Jeremiah Ostriker, and Guohong Xu for supplying the
output of their numerical simulation and Joachim Stadel and the
University of Washington's NASA HPCC ESS group for making public their
very efficient SMOOTH and FOF algorithms
(http://www-hpcc.astro.washington.edu/tools).  DJE was supported by
NSF grant PHY-9513835.
The source code for HOP is available at 
http://www.sns.ias.edu/$\sim$eisenste/hop/hop.html.

\clearpage
\centerline{\bf Figure Captions}
\bigskip

\noindent{\bf Figure 1:} An illustration of the algorithm to merge maxima.
The shaded regions represent areas with density exceeding $\dm$.
The solid heavy contours represent $\dout$, the dashed heavy
contours represent $\ds$, and light contours represent other values.
Maxima A and B are merged into a single group because they lie within
a single $\ds$ contour; C and E are separate groups because they do
not share this contour.  The maxima D, F, G, and H cannot be group centers
because they do not achieve $\delta=\dm$; they are attached to other
groups because they share $\dout$ contours with them.  F, G, and H are
attached to E, while D is attached to the A--B group rather than C
because it shares a higher boundary with the former.  The situation with
F and G demonstrates the benefit of having $\ds<\dm$; had they been
equal, this small fluctuation in the contour would have split the group.
All groups would be truncated at density $\dout$, resulting in the
three groups enclosed by the solid and dashed-dot heavy contours.

\noindent{\bf Figure 2:} Plot of the fraction of mass in groups per
logarithmic interval of group mass versus the logarithm of the group
mass.  Our canonical case is $\nd=64$, $\nh=16$, $\nm=4$, $\dm=120$,
and $\dout=80$.  Here we show three curves, varying $\nh$ from 8 to 64.

\noindent{\bf Figure 3:} Same as Figure 2, but $\nm$ is varied from 
1 to 8.

\noindent{\bf Figure 4:} Same as Figure 2, but $\nd$ is varied from 64
to 128 and also compared to the results using $\nd=24$ with a top-hat
kernel rather than the cubic spline.

\noindent{\bf Figure 5:} Same as Figure 2, but $\dm$ and $\ds$ are varied
at fixed $\dout$.  We maintain $\ds=(3\dm+\dout)/4$.

\noindent{\bf Figure 6:} Same as Figure 2, but $\ds$ is varies from
80 to 240 while $\dout$ and $\dm$ are held constant at 80 and 240,
respectively.

\noindent{\bf Figure 7:} Same as Figure 2, but we vary the density
cutoffs while holding $\dm=2\dout$ and $\ds=1.75\dout$.  Also shown are
the results for two choices of friends-of-friends linking parameter
$b$, in units of the mean interparticle spacing.  The top pair of
curves are $\dout=600$ and $b=0.1$ and the bottom pair are $\dout=80$
and $b=0.2$.

\clearpage
\centerline{\epsfxsize=\textwidth \epsffile{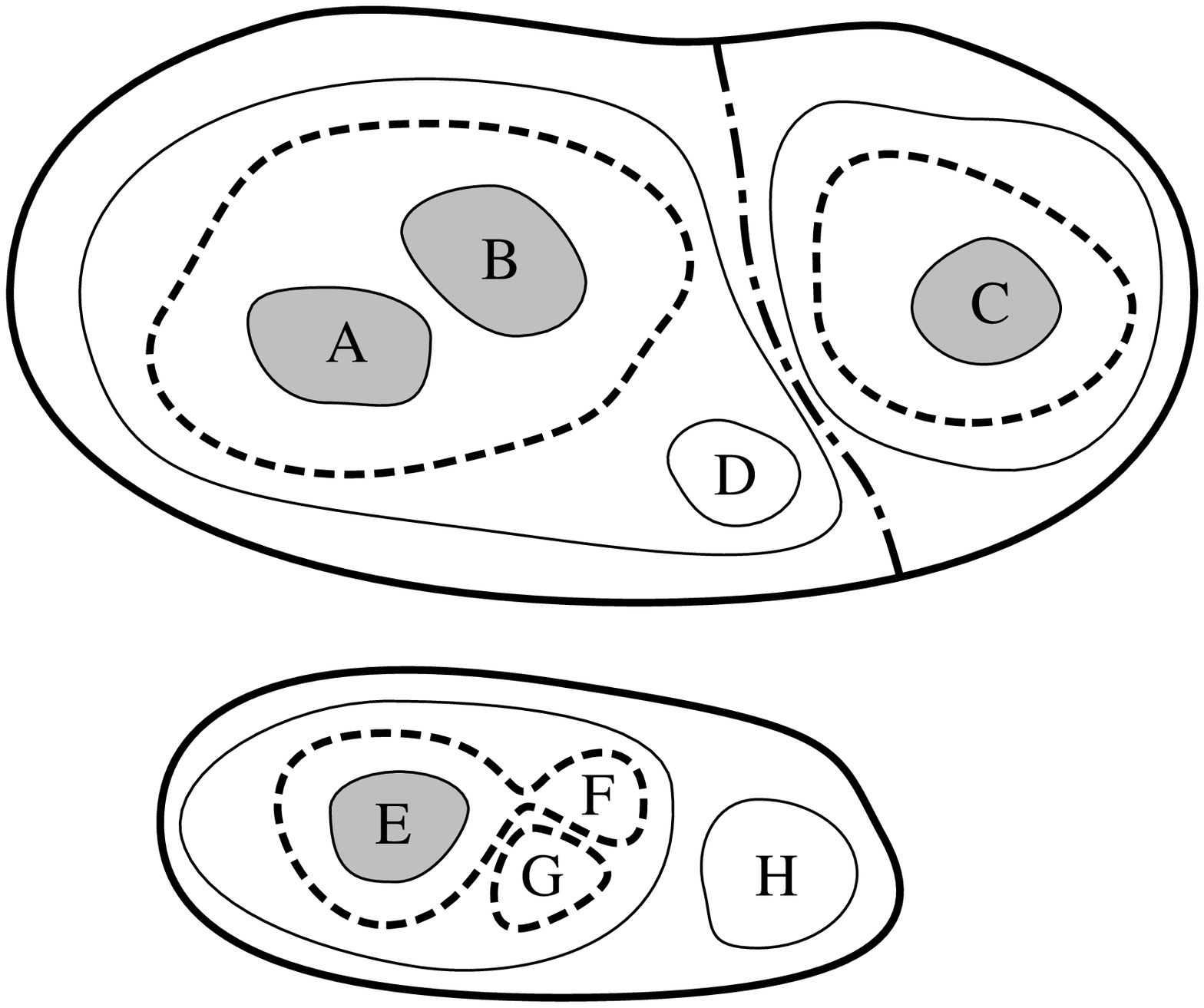}}
\clearpage
\centerline{\epsfxsize=\textwidth \epsffile{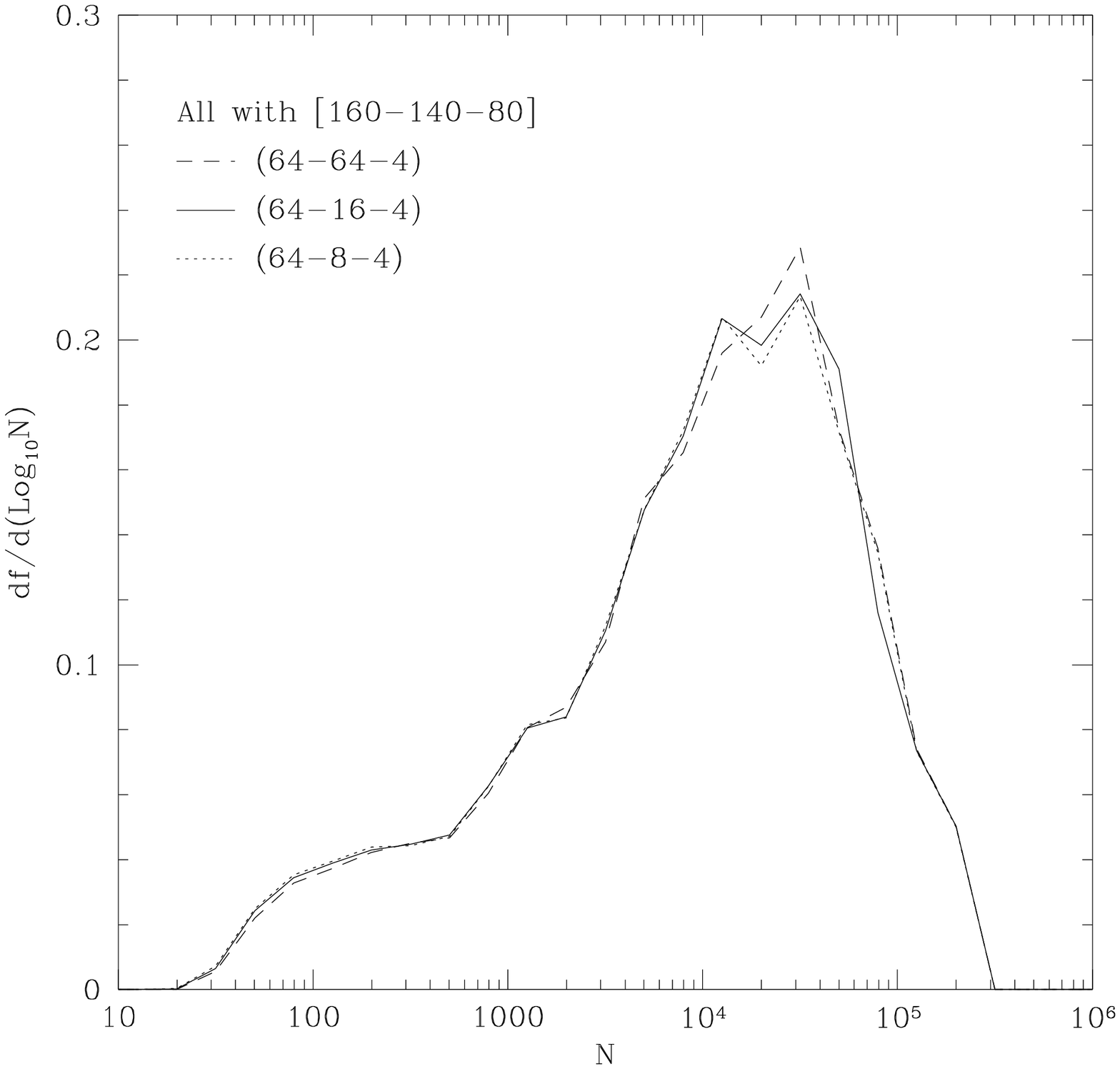}}
\clearpage
\centerline{\epsfxsize=\textwidth \epsffile{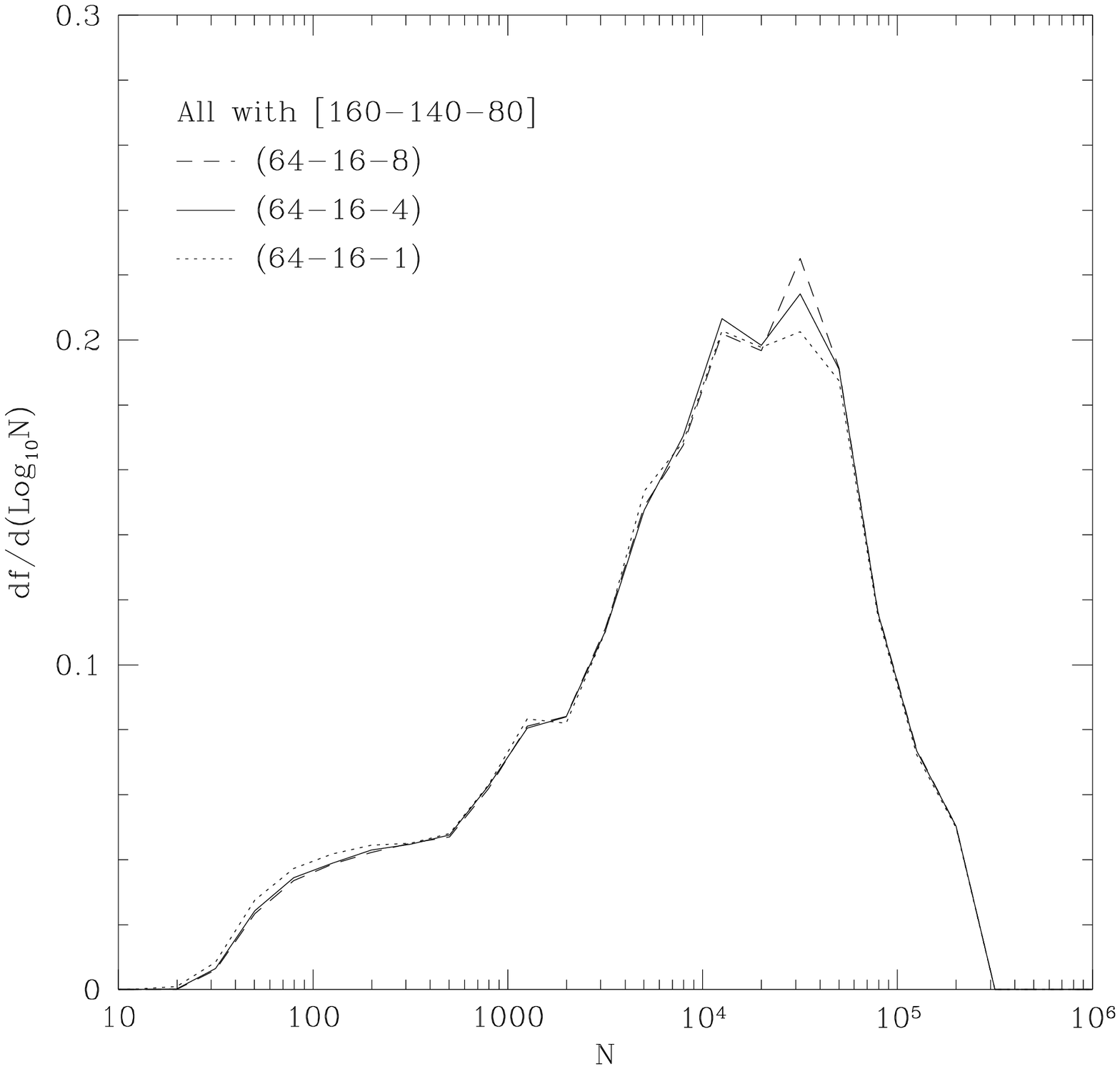}}
\clearpage
\centerline{\epsfxsize=\textwidth \epsffile{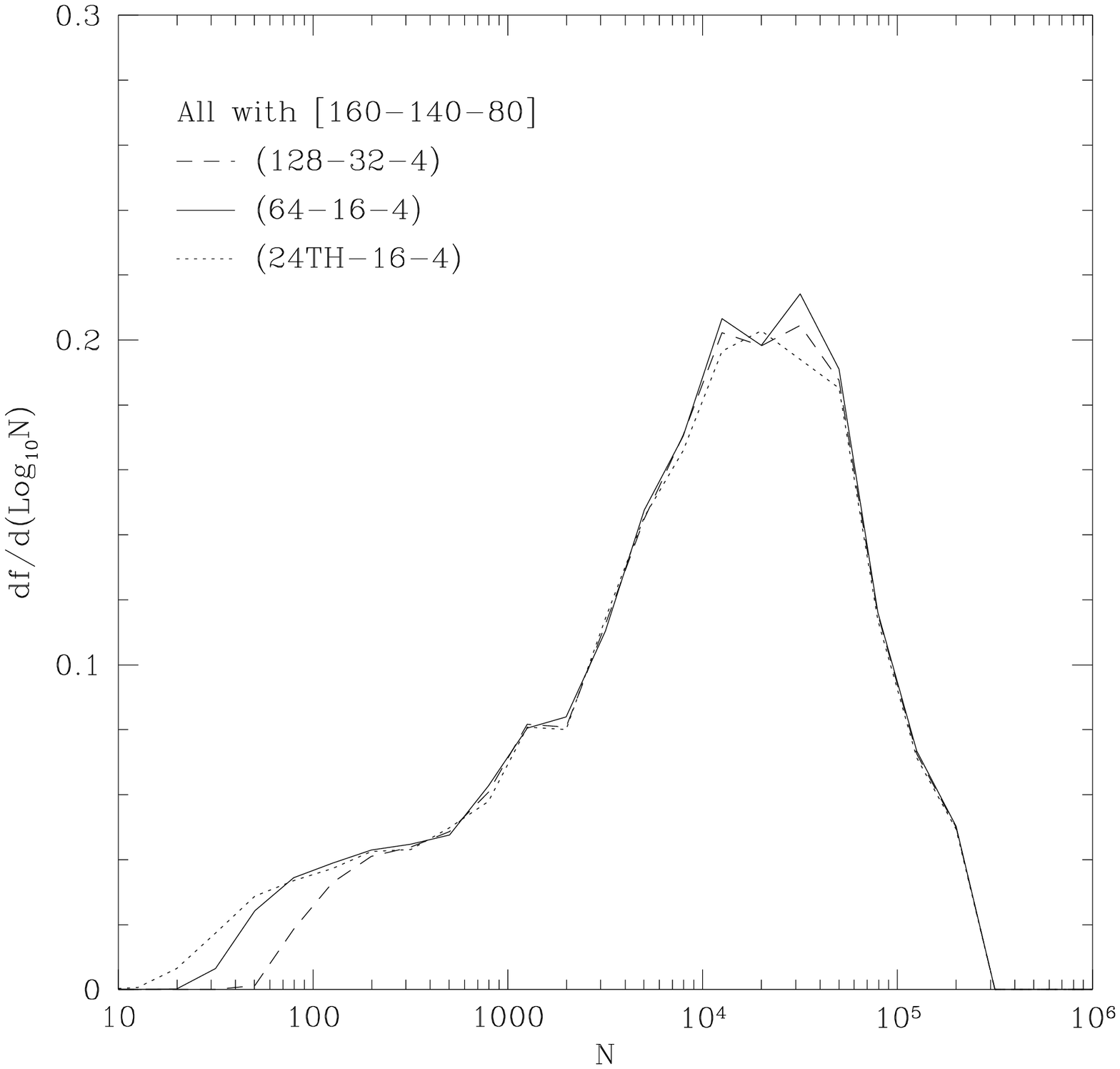}}
\clearpage
\centerline{\epsfxsize=\textwidth \epsffile{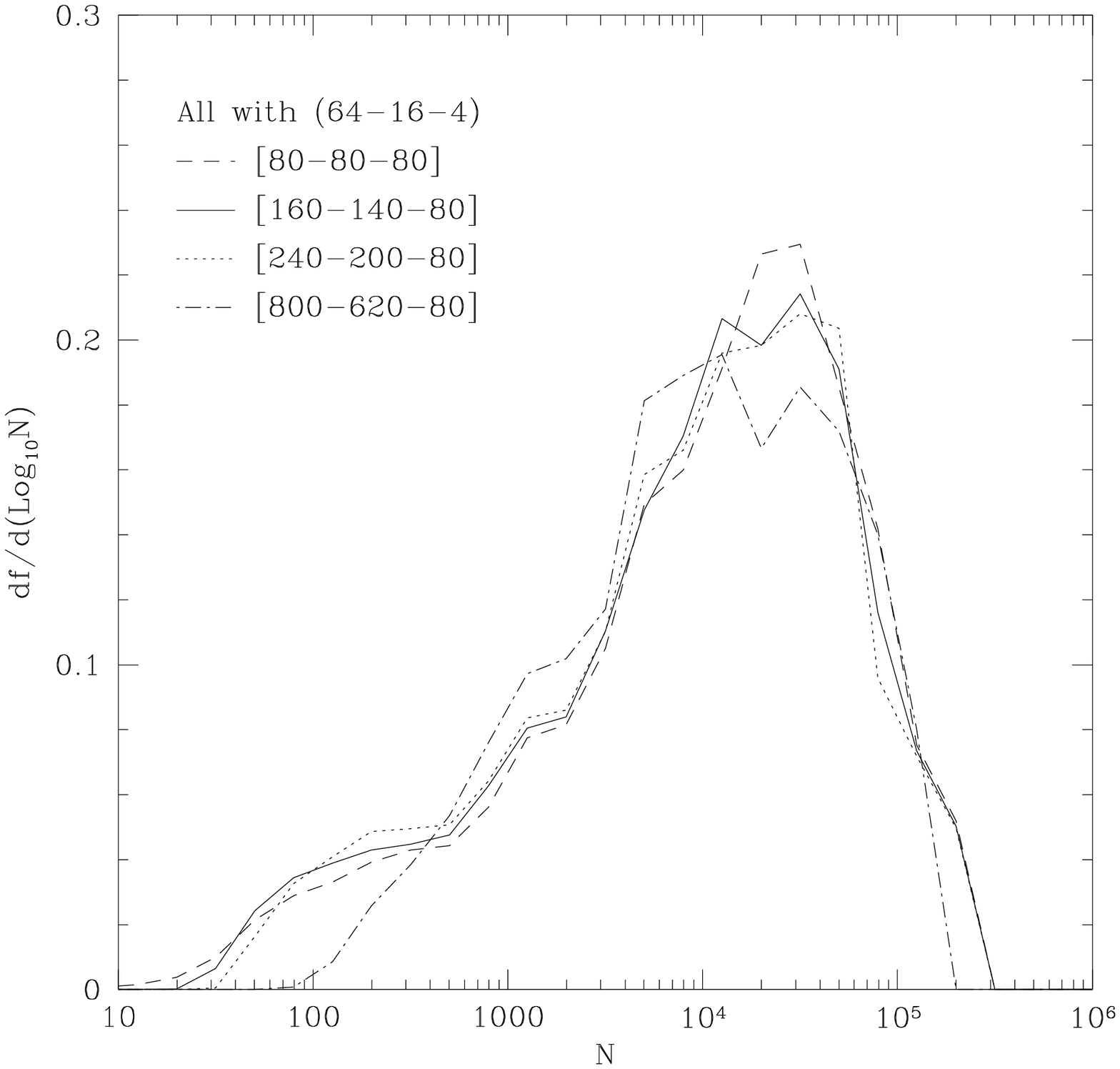}}
\clearpage
\centerline{\epsfxsize=\textwidth \epsffile{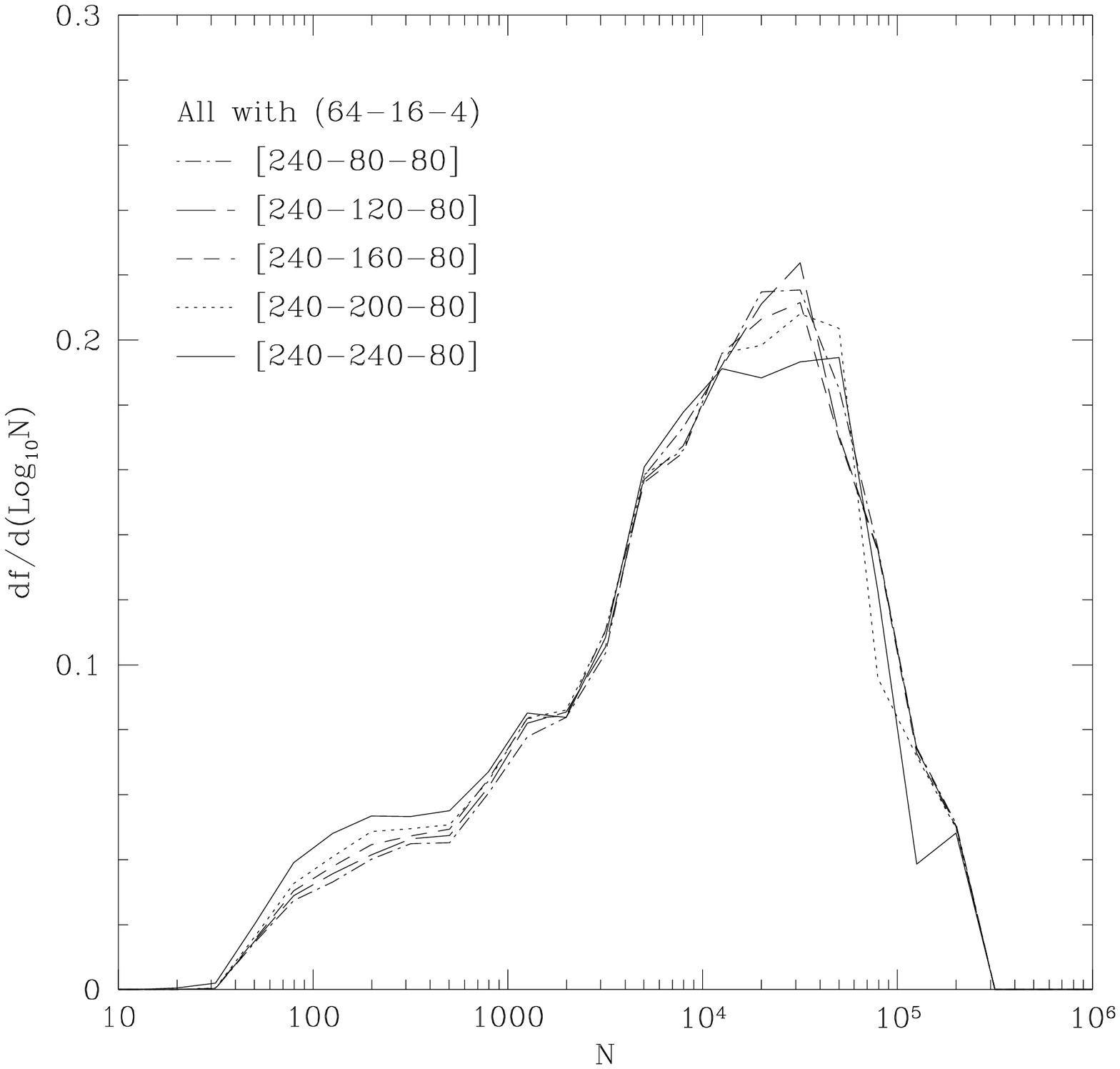}}
\clearpage
\centerline{\epsfxsize=\textwidth \epsffile{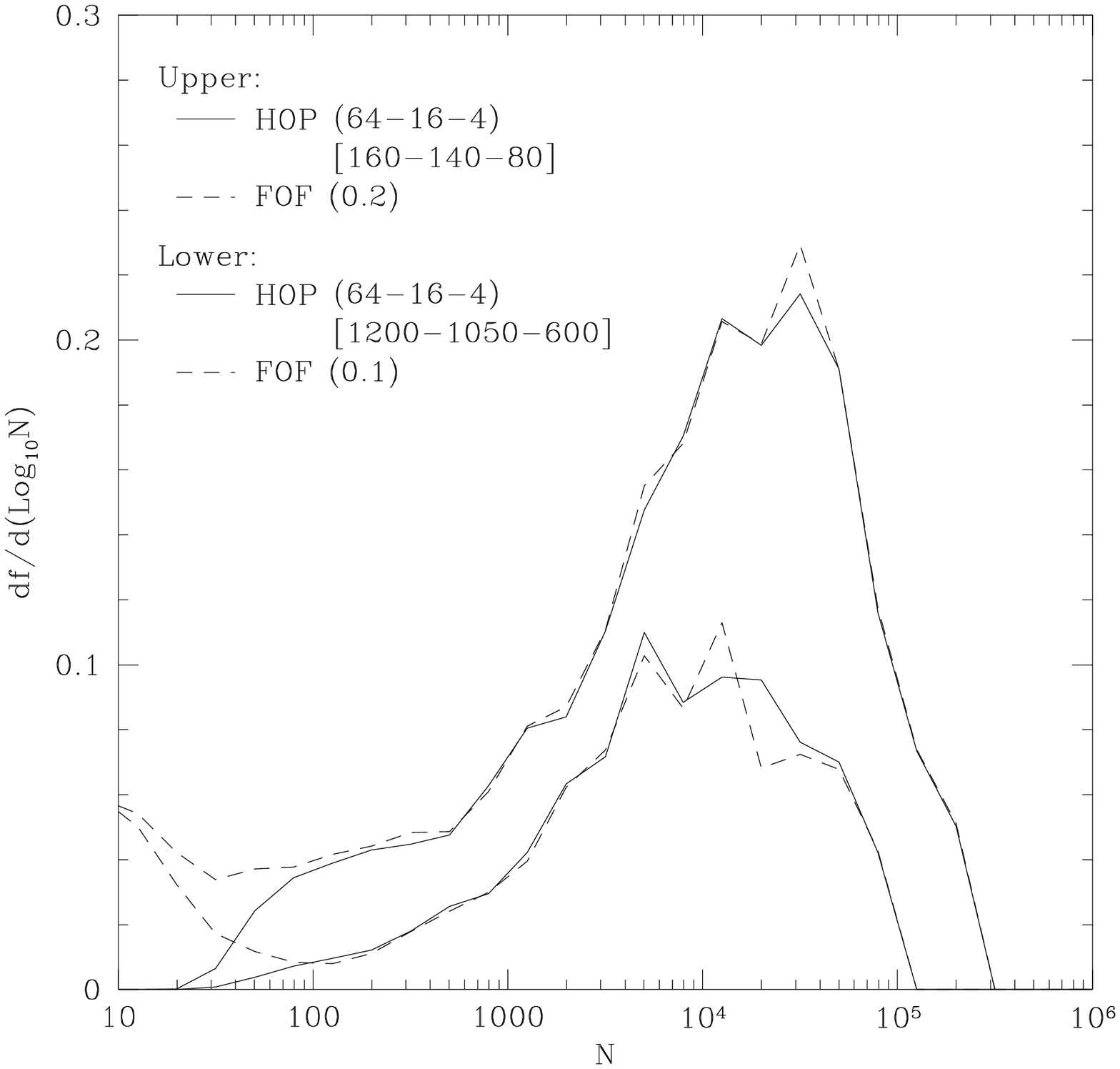}}

\end{document}